\documentclass[a4paper]{jpconf}

\usepackage{graphicx}

\usepackage{amsmath}

\usepackage{mathrsfs}%,euler}

\usepackage{comment}

\usepackage{color}

\usepackage{amssymb}

\usepackage{amsthm}

\usepackage{graphicx}

\usepackage{upgreek}

\usepackage{mathtools}

\usepackage{empheq}

\usepackage{amsmath,amssymb,amsthm,mathtools}

\usepackage{listings} 

\usepackage{braket}

\usepackage{tikz}

\usepackage[toc,page]{appendix}

\usetikzlibrary{arrows,shapes,calc}

\usepackage{mathpazo}

\usepackage{cite}

\usepackage{palatino}

\usepackage{wrapfig}

\usepackage{epsf,color,colordvi,pifont}

\usepackage{nicefrac}

\DeclareFontFamily{OT1}{pzc}{}
\DeclareFontShape{OT1}{pzc}{m}{it}{<-> s * [1.10] pzcmi7t}{}
\DeclareMathAlphabet{\mathpzc}{OT1}{pzc}{m}{it}

\begin{document}
\title{Axion Electrodynamics in Strong Magnetic Backgrounds}

\author{S  Villalba-Ch\'avez$^1$, A  E Shabad$^{2,3}$, C  M\"uller$^1$}

\address{$1$ Institut f\"ur Theoretische Physik I, Heinrich-Heine-Universit\"at D\"usseldorf, Universit\"atsstr. 1, 40225  D\"usseldorf, Germany}

\address{$2$ Lebedev Physical Institute, 53, Leninsky prospect, 119991, Moscow, Russia}

\address{$3$ Tomsk State University, 36, Lenin Prospekt, 634050, Tomsk, Russia}

\ead{villalba@uni-duesseldorf.de}

\begin{abstract}
The overcritical regime of axion-electrodynamics (AED) is investigated. For magnetic fields larger than the characteristic scale linked to AED, quantum vacuum fluctuations due to axion-like fields can dominate over those associated with the electron-positron fields. This hypothetical regime of the dominance of AED over QED is predicted to induce  strong  birefringence  and screening effects. We show that,  if  the  magnetic field lines are curved,   extraordinary photons could be  canalized  along the field direction.  It is also shown that  the running QED coupling depends on the magnetic field strength, and for certain energy regimes, it could be screened almost to zero, making the QED building blocks very weakly interacting between each other. The impact of this phenomenon on the radiation mechanism of pulsars is discussed.
\end{abstract}

%%%%%%%%%%%%%%%%%%%%%%%%%%%%%%%%%%%%%%%%%%%%%%%%%%%%%%%%%
\section{Introduction}
%%%%%%%%%%%%%%%%%%%%%%%%%%%%%%%%%%%%%%%%%%%%%%%%%%%%%%%%%

The  Peccei-Quinn mechanism  is  among the  most appealing solutions to the  so-called strong CP problem. However, it predicts the existence of a pseudoscalar Nambu-Goldstone boson which has not been observed so far: the axion  \cite{Peccei:1977hh,Wilczek:1977pj,Weinberg:1977ma}. Constraints on the invisible axion \cite{Donnelly,Zehnder,Dine:1981rt,zhitnitskii,kim,shifman}-- or more general -- axion-like particles  are deduced from their potential astro-cosmological consequences which are not reflected accordingly by the current observational data of our  universe  \cite{Jaeckel:2010ni,Ringwald:2012hr,Alekhin:2015byh,Javier}.  A basic assumption underlying this line of argument is that the interplay between  axion-like particles and the well established Standard Model  sector --  photons in first place -- is extremely feeble. As a consequence, axion-like particles produced copiously in the core of stars via the Primakoff effect might escape from there almost freely, constituting a leak of energy that accelerates the cooling of the star and, thus, shortens its lifetime. Therefore, the number of red giants in  the helium-burning phase in globular clusters should diminish considerably. That this fact does not take place -- at least not significantly -- constraints the axion-diphoton  coupling $g$ to lie below $g\lesssim 10^{-10}\ \mathrm{GeV^{-1}}$ for axion masses $m$ below the $\mathrm{keV}$ scale \cite{Raffelt:1985nk,Raffelt:1999tx,Raffelt:2006cw}.

Precisely on the surface of stellar objects identified as neutron stars \cite{Manchester,Kouveliotou,Alaa,Bloom} and magnetars, magnetic fields as large as $B\sim \mathcal{O}(10^{13}-10^{15})\ \mathrm{G}$  are predicted to exist. As these strengths are bigger than the characteristic QED scale $B_{0}=4.42\times 10^{13}\ \mathrm{G}$  \cite{Berestetsky,Schwartz,Dittrich}, such astrophysical scenarios are propitious for  the realization of a variety of yet unobserved quantum processes, which are central in our current understanding of the nature and origin of the pulsar radiation. Notable among them are: the photon  capture effect by the magnetic field owing to the resonant behavior \cite{shabad1972} of the vacuum polarization in QED \cite{shabadnat,shabad3v} and  the photon splitting effect \cite{adler1,adler2,adler3}. Clearly, the strong-field environments provided by these compact objects can also be favorable for an axion-like particles phenomenology -- as they are in QED --  contrary to what is predicted relying on the weak coupling assumption. This occurs, because the aforementioned field strengths could compensate for the weakness of the coupling and significantly   stimulate  quantum vacuum fluctuations of axion-like fields.  

Theoretically, the problem  linked to field strengths overpassing  the corresponding  electric $E_c=m/g$ and magnetic $B_c=m/g$  scales of AED have not yet attracted much attention, save in \cite{Mateja,Villalba}, where their respective relevance  has  been analyzed  in connection with phenomenologies  predicted to take place in  black holes and neutron stars.  While   an  electric  background $E\gtrsim E_c$  induces an axionic instability,  for magnetic fields $B\gtrsim B_c$  the theory is stable  and -- as in QED -- the screening and refractive properties  induced by  quantum vacuum fluctuations of  axion-like fields are predicted to be more pronounced. 

%%%%%%%%%%%%%%%%%%%%%%%%%%%%%%%%%%%%%%%%%%%%%%%%%%%%%%%%%%%%%%%%%%
\section{Refraction properties in a constant magnetic field    \label{Sec:AIRACOGQ}}
%%%%%%%%%%%%%%%%%%%%%%%%%%%%%%%%%%%%%%%%%%%%%%%%%%%%%%%%%%%%%%%%%%

\begin{wrapfigure}{l}{0.45\textwidth}
 \includegraphics[width=0.45 \textwidth]{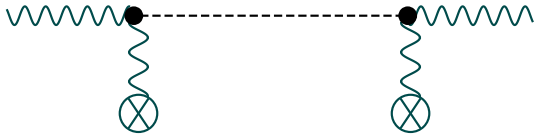}
\caption{Diagrammatic representation  of   the  magnetic-field dependent part of the vacuum polarization tensor mediated by quantum vacuum fluctuation of an axion-like field.}
\label{fig:001}
\vspace{-10pt}
\end{wrapfigure}

\begin{wrapfigure}{r}{0.45\textwidth}\centering
 \includegraphics[width=0.45\textwidth]{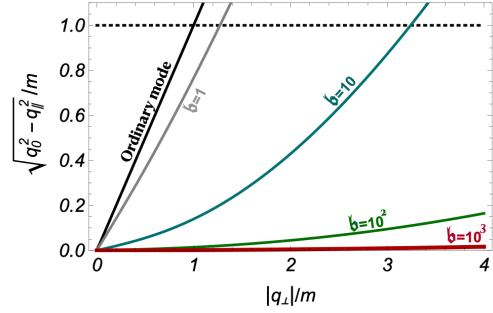}
\caption{Dispersions curves for the ordinary (in black) and the extraordinary mode (gray, cyan, green and red)  [see Eq.~(\protect\ref{dispersionrelations})].}
\label{fig:003}
\vspace{-60pt}
\end{wrapfigure}

When  the Lagrangian density  of AED  is embedded in the  QED action and a magnetic field background is considered,  quantum vacuum fluctuations of the axion-like field introduce a contribution to the polarization tensor, as shown  in Fig.~\ref{fig:001}.\footnote{Also  a   one-loop  contribution is involved \cite{alina}. However, in a strong field $B\gg B_c$ it can be ignored.} For magnetic fields larger than\footnote{We use units with the  Planck constant  $\hbar$, the speed of light $c$  and the vacuum permittivity $\epsilon_0$  taken to be  unity [$\hbar=c=\epsilon_0=1$].}  
  the  characteristic scale [$B_c=\mathpzc{m}/g$]  linked to  AED,  the phenomenology of the quantum vacuum could be ruled by virtual ALPs instead of  fluctuations of the electron-positron field. This  seems to be particularly  accessible because the  vacuum polarization tensor $\Pi^{\mu\nu}(x,\tilde{x})$ in AED  exhibits a quadratic growth in the external  field while  the  corresponding  QED polarization tensor grows linearly  \cite{melrose,heyl}, provided $B\gg B_0$ with  $B_0=m_e^2/e\approx4.42\times 10^{13}\ \rm G$ referring to the characteristic scale in QED. The consequence of this  conjecture of axion dominance  is  explored in this contribution. The starting point of our analysis is the  equation of motion for a small-amplitude electromagnetic wave $a_\mu(x)$  with $\partial a=0$,  modified by $\Pi^{\mu\nu}(x,\tilde{x})$ 
\begin{equation}\label{DE}
\square a^\mu(x)+i\int d^4\tilde{x}\ \Pi^{\mu}_{\ \nu}(x,\tilde{x})a^{\nu}(\tilde{x})=0.
\end{equation} Its solutions in  the  Fourier space are characterized by a massive mode and two massless states. Among the massless excitations, there is an ordinary mode with dispersion relation $q_{0,o}=\vert\pmb{q}\vert$ and  an extraordinary mode, the dispersion law of which reads [see Fig. \ref{fig:003}]
\begin{equation}
q_{0,e }^{2}=\pmb{q}^{2}+\frac{1}{2}m_{\ast }^{2}\left[ 1-\sqrt{1+\frac{4%
\mathfrak{b}^{2}}{1+\mathfrak{b}^{2}}\frac{q_{\perp }^{2}}{m_{\ast }^{2}}}%
\right],
\label{dispersionrelations}
\end{equation}  where $m_*=m\sqrt{1+\mathfrak{b}^2}$ is the ALP mass dressed by the magnetic field  parameter $\mathfrak{b}=B/B_c$.  Here $q_\perp$ ($q_\parallel$) is the momentum perpendicular (parallel) to the field direction and $\vert\pmb{q}\vert ^2=q_\parallel^2+q_\perp^2$. 

In contrast to the ordinary mode, the dispersion relation of the extraordinary one  depends on the magnetic field strength which reveals the axion-modified birefringent feature of the vacuum.  When calculating  the components of the group velocity   $\mathpzc{v}_{\perp,\parallel}=\partial q_{0}/\partial q_{\perp,\parallel}$ linked to  the extraordinary branch,  we find  [$q_\perp\ll m_*$ ]
\begin{equation}
\begin{split}
&\mathpzc{v}_{\parallel,e}\approx \frac{q_\parallel}{q_{0,e}},\\
&\mathpzc{v}_{\perp,e}\approx \frac{1}{1+\mathfrak{b}^2}\frac{q_\perp}{q_{0,e}}\left[1+\frac{2\mathfrak{b}^4}{1+\mathfrak{b}^2}\frac{q_\perp^2}{m_*^2}\right]. 
\end{split} \label{q-par-is-zero}
\end{equation}As a consequence, the angle  between the direction of  propagation of the electromagnetic energy and the external magnetic field  $\theta=\arctan(\mathpzc{v}_\perp/\mathpzc{v}_\parallel)$ does not coincide with the one  between the photon  momentum $ \pmb{q}$  and  $\pmb{B}$, i.e. $\vartheta=\arctan(q_\perp/q_\parallel)$.  Furthermore, we  observe that it tends to vanish when  $q_{\perp }\ll m_*\approx m\mathfrak{b}$ the faster, the stronger the field is, since one has $ \mathpzc{v}_{\perp,e}\to 0$ and $\mathpzc{v}_{\parallel,e}\to1$. This implies that the group velocity of extraordinary excitations tends to be  parallel to the magnetic field  for hard, as well as for soft photons with $q_\perp\ll m_*$. Beyond this limit, i.e. for $q_\perp\gg m_*$, the components of the group velocity approach $\mathpzc{v}_{\perp,e}\approx q_\perp/q_{0,e}$,  $\mathpzc{v}_{\parallel,e}=q_\parallel/q_{0,e}$, and the angle $\theta$ between the direction of propagation of the electromagnetic energy and $\pmb{B}$ does not deviate substantially  from the one formed by the momentum of the small-amplitude wave and the external field [$\theta\approx \vartheta$]. 

%%%%%%%%%%%%%%%%%%%%%%%%%%%%%%%%%%%%%%%%%%%%%%%%%%%%%%%%%%%%%%%%%%
\section{Running  QED  coupling  and modified Coulomb potential \label{Sec:RQEDCAMCP}}
%%%%%%%%%%%%%%%%%%%%%%%%%%%%%%%%%%%%%%%%%%%%%%%%%%%%%%%%%%%%%%%%%%

The problem of determining how the quantum vacuum fluctuation of a pseudoscalar axion field  modifies the Coulomb potential $a_C(\pmb{x})=\mathpzc{q}/(4\pi \vert\pmb{x}\vert)$ reduces, to a large extent, to determine  the explicit expression for the photon Green function including the vacuum polarization effect. This can be reached by inverting the operator which defines the equation of motion [see  Eq.~\eqref{DE}].  As a consequence,   the following expression for the  axion-modified Coulomb potential was found [for details we refer the reader to Ref.~\cite{Villalba}]:
\begin{equation}
\begin{split}
a_0(\pmb{x})=\mathpzc{q}\int  \frac{d^3q}{(2\pi)^3}\frac{e^{-i\pmb{q}\cdot\pmb{x}}}{\pmb{q}^2}\frac{ 1}{1+\frac{g^2B^2q_\parallel^2}{\pmb{q}^2+m^2}},
\end{split}
\label{initialpotential}
\end{equation}where $\mathpzc{q}$ is the charge of a point-like source.

Let us consider the electrostatic energy $\mathpzc{U}(\pmb{x})=-e a_0(\pmb{x})$ of  an electron. The insertion of Eq.~(\ref{initialpotential})  into this formula  with $\mathpzc{q}\to e$ determines how  quantum vacuum fluctuations of axion-like fields in the presence of a strong magnetic background   [$\mathfrak{b}\gg1$]  modify  the running QED coupling  
\begin{equation}
\alpha_{\mathrm{EM}}(q_\parallel^2,q_\perp^2)\approx\frac{\alpha}{1+\frac{g^2B^2q_\parallel^2}{\pmb{q}^2(\pmb{q}^2+m^2)}}
\label{runningcoupling}
\end{equation} with  $\alpha=e^2/(4\pi)\approx137^{-1}$  defined  at $\mathfrak{b}\to0$ and   $\vert\pmb{q}\vert\to0$ \cite{Berestetsky,Schwartz}.   Notably, if the dominance of AED over QED takes place,  $\alpha_{\mathrm{EM}}(\pmb{q})$  is free of  the characteristic Landau pole arising  in  QED  that rises  the coupling constant  to an infinite value for large enough momentum. Furthermore, its dependence on the  magnetic field  leads to an anisotropic behavior  along and transverse to $\pmb{B}$. This dependence  decreases the value of  the running QED coupling  as the magnetic field $B$  grows  while  keeping the momentum fixed. Indeed, for $q_\perp^2\ll q_{\parallel}^2\ll m^2$,  one finds that $\alpha_{\mathrm{EM}}(q_\parallel^2,q_\perp^2)\sim \alpha/\mathfrak{b}^{2}\ll1$  tends to be screened to zero. This phenomenon makes  QED  an almost ``trivial''  theory in the aforementioned  regime, meaning that the corresponding charged  matter sector and the  small-amplitude electromagnetic waves turn out to be very weakly interacting between each other. 
\begin{figure*}[t]
\centering
\includegraphics[width=6in]{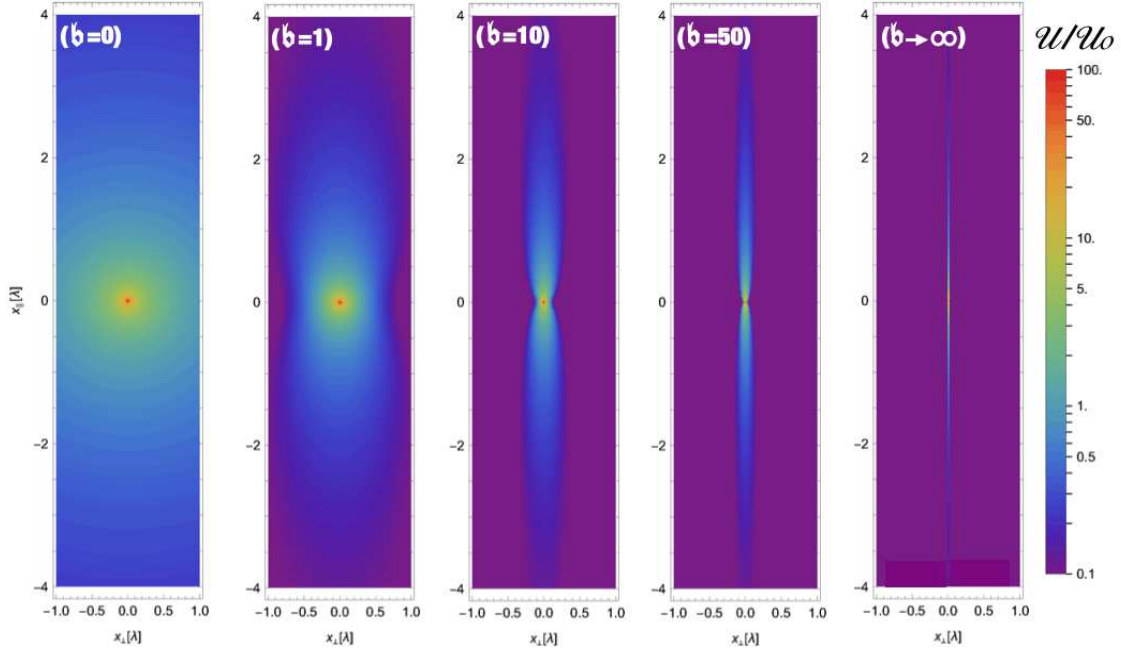}%
\caption{Electrostatic energy of an electron   $\mathpzc{\mathpzc{U}(\pmb{x})}=-ea_0(\pmb{x})$ due to the anisotropic Coulomb potential $a_0(\pmb{x})$ of a  point-like static charge surrounded by quantum vacuum fluctuations of an axion-like field over a constant magnetic-like background. Here the energy is shown in units of the electron energy at the Compton wavelength of an axion  $\mathpzc{U}_0=-e\mathpzc{q}/(4\pi \lambda)$ with $\lambda=2\pi m^{-1}$. }
\label{fig:mb002}
\end{figure*}
This situation becomes particularly dramatic if the axion mass turns out to be larger than  the first pair creation  threshold, i.e. $m> 2m_e$, where $m_e$ refers to  the electron mass.  Indeed, under such a restriction,  radiation channels such as the  recombination of pairs, cyclotron radiation and  the photon splitting effect  along with    the main pair production mechanisms  in highly-magnetized pulsars would  be suppressed.  This  sort of quantum triviality \cite{Holger,Landau,Efrain}  persists  in domains other than the one  discussed before \cite{Villalba}. Conversely,    for $q_{\perp}^2/m^2\ll 1\ll \mathfrak{b}^2 \ll  q_\parallel^2/m^2$ or $1\ll \mathfrak{b}^2\ll q_\perp^2/q_\parallel^2\ll m^2/q_\parallel^2$ the screening effects are negligibly small and the running QED coupling approximates to its standard value $\alpha_{\mathrm{EM}}(q_\parallel^2,q_\perp^2)\sim \alpha$.  

The   described  properties  of   the  running  electromagnetic coupling have direct implications on  the Coulomb potential  [see Eq.~\eqref{initialpotential}]. For strong magnetic fields  [$\mathfrak{b}\gg1$] and at distances from the source  much smaller than the Compton wavelength of the axion  $2\pi m^{-1}\gg \vert\pmb{x}_{\perp,\parallel}\vert$, the Coulomb potential of a point-like charge turns out to be of  Yukawa-type  in the direction perpendicular to the magnetic field while along $\pmb{B}$ it follows approximately half of the Coulomb law:
\begin{equation}
a_0(\pmb{x}_\perp,0)\approx \frac{\mathpzc{q}}{4\pi\vert\pmb{x}_\perp\vert}e^{-\frac{1}{2}m_*\vert\pmb{x}_\perp\vert},\quad  a_0(0,\pmb{x}_\parallel)\approx\frac{1}{2} \frac{\mathpzc{q}}{4\pi\vert\pmb{x}_\parallel\vert}  \label{xperp}
\end{equation} 

Figure~\ref{fig:mb002}  exhibits the evolution of the  electrostatic energy of an electron $\mathpzc{U}(\pmb{x})=-ea_0(\pmb{x})$ as the magnetic field grows gradually. This evaluation was carried  out from  Eq.~\eqref{initialpotential}. While  the  panel  most to the left has been obtained by considering  the standard Coulomb potential $a_C(\pmb{x})=\mathpzc{q}/(4\pi\vert\pmb{x}\vert)$, the remaining ones were generated by taking the field parameter $\mathfrak{b}=1$, $\mathfrak{b}=10$, $\mathfrak{b}=50$ and  $\mathfrak{b}\to\infty$, respectively. These results show that, as the field  strength  increases, the static field of an electric charge is anisotropic and  tends to be  squeezed towards the   axis   parallel to $\pmb{B}$. This behavior is a  clear manifestation of a dimensional reduction  linked to the  corresponding gauge sector,\footnote{Only  virtual  modes with  the polarization of the extraordinary  photons carry the electrostatic interaction.} and it still appears  at distances  for which  $\vert\pmb{x}_{\perp,\parallel}\vert\gg m_{*}^{-1}$, where the Coulomb's law  acquires the form
\begin{equation}
\begin{split}
&a_0(\pmb{x})\approx\frac{\mathpzc{q}}{4\pi\sqrt{\vert\pmb{x}_\perp\vert^2(1+\mathfrak{b}^2)+\vert\pmb{x}_\parallel\vert^2}}.
\end{split}
\label{strongfieldlargexparallel2}
\end{equation} Unless $\vert\pmb{x}_\perp\vert=0$, this formula  tends to vanish as the external field grows and $\vert \pmb{x}_\perp\vert\gg \mathfrak{b}^{-1}\vert \pmb{x}_\parallel\vert$. Conversely, for $\mathfrak{b}\gg1$ and $\vert \pmb{x}_\parallel\vert\gg \mathfrak{b}\vert \pmb{x}_\perp\vert$, the Coulomb potential tends to be one-dimensional $a_0(\pmb{x})\approx \mathpzc{q}/(4\pi\vert\pmb{x}_\parallel\vert)$. We remark that the expression above  resembles closely the longe-range behavior  found within a pure QED context, when the external magnetic field exceeds the characteristic scale associated with this framework $B_0=m_e^2/e\approx4.42\times 10^{13}\ \rm G$ 
\cite{shabad5,shabad6,Adorno2016}. 

%%%%%%%%%%%%%%%%%%%%%%%%%%%%
\section{Captures of $\gamma$ quanta: implications }
%%%%%%%%%%%%%%%%%%%%%%%%%%%%
 
Let us suppose  a magnetic field with its lines of force curved as it is adopted in models describing the magnetosphere of pulsars. The fields of these  highly magnetized  objects  rotate  with certain angular velocity $\Omega$. However,  as long as both its  curvature radius  and its  rotation period $2\pi\Omega^{-1}$   exceed considerably  the photon wavelength and the Compton wavelength of  ALPs, the analysis developed above   can still  be applied.  Indeed, whenever the previous requirements are fulfilled   a  geometrical optical and an adiabatic  approximation can be used simultaneously  to investigate the propagation of those photons subject to the aforementioned limitations.    To accomplish them,  we shall refer hereafter to the so-called curvature $\gamma $-radiation  ($q_{0}>2m_{e}$) which is  emitted tangentially to the magnetic field lines.  At the moment of its emission this electromagnetic wave  decomposes into  ordinary and extraordinary modes and  both of them are localized   at the origin  of the diagram in Fig.~\ref{fig:003}.  As the wave  propagates,  the   momentum components $q_{\perp}$ and $q_{\parallel }$ of theses modes  change, while their energies $q_{0}$ are  kept constant approximately.   Thereby the dispersion relation  of the ordinary and  extraordinary photons  evolve  from the moment of their emission following  the corresponding   curves displayed in  Fig.~\ref{fig:003}. 

\begin{wrapfigure}{r}{0.45\textwidth}\centering\vspace{-15pt}
 \includegraphics[width=0.45\textwidth]{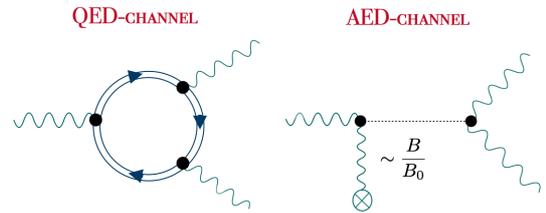}\vspace{-20pt}
\caption{Photon splitting in QED$+$AED. A double-line  in the  QED diagram represents the electron-positron propagator including  the interaction  with the external field.  The dashed line in the AED-channel denotes the free axion propagator.}
\label{fig:003sp}
\vspace{-10pt}
\end{wrapfigure}

Since   the curvature radiation appears well below the  first pair creation threshold, the main  process  taking place in this regime is the  photon splitting: $\gamma+B\to \gamma^\prime+\gamma^{\prime\prime}+B$. This  inelastic scattering  becomes feasible  as long  as  the photon wave vector $\pmb{q}$ and the magnetic field $\pmb{B}$ form a nontrivial  angle [$\vartheta\neq 0$].  Under such a circumstance  the  ordinary mode  is   expected to decay into two extraordinary photons before reaching  the first pair creation threshold,  provided $B\gtrsim 0.2\; B_0$ \cite{shabad3v}. Mainly because the  electromagnetic sector linked to this type of degree of freedom  is unaffected by ALPs, and therefore,  it is not subject to the aforementioned dominance.  As a result of the  decay of the ordinary mode,  the curvature radiation acquires  a high  degree of polarization   in the plane spanned by $\pmb{q}$ and $\pmb{B}$ before the extraordinary modes  reach the first pair creation threshold. 

The axion-diphoton interaction  opens a new channel for the splitting of a photon [see Fig.~\ref{fig:003sp}].  Observe that,  owing to the parity invariance,  only ordinary $\gamma$ photons can decay via the AED channel, and that  the result of the reaction involves  photons with different nature.  We remark that, in the equation of motion,  the  term linked to this  photon splitting  turns out to be  sub-leading as compared  to the term involving  the axion-mediated polarization tensor. Mainly  because  it depends  linearly on the external field. 

Now, for $q_{\perp}\ll m_*$  the   dispersion law of an extraordinary  mode  tends to stick to the horizontal axis  as $\mathfrak{b}$ grows. This is precisely   the region in which  the group velocity  across the magnetic field disappears [$v_{\perp,e}\to 0$].  Hence,  this  $\gamma$ quantum would be   canalized  along the magnetic line of force   immediately  after  its emission.  It is worth remarking that,  if this photon capture occurs,  a single sufficiently hard -- belonging to the $\gamma $-range -- photon  might  have no  chance to produce  an electron-positron  pair  before it  has propagated far enough  to leave the region of the  strong dipole field  of the pulsar where, however, the one-photon pair creation is strongly suppressed.  This is guaranteed if the distance  traversed  by the extraordinary photon is such that the local magnetic field satisfies the relation [$b=B/B_0$]   
\begin{equation}\label{initialrelation}
b\ll1\ll\mathfrak{b}
\end{equation}and, up to this point the lowest pair creation threshold  is not reached i.e. $q_{0.e}^2-q_\parallel^2<4m_e^2$. While the former relation introduces the condition $B_c\ll B_0$ along with $\mathfrak{b}\gg1$, the latter provides a characteristic value for  $q_\perp$ \cite{Villalba}:
\begin{equation}
(q_{\perp,\mathrm{thr}})^2=4m_e^2+\frac{1}{2}m^2\left(\sqrt{1+16\mathfrak{b}^2\frac{m_e^2}{m^2}}-1\right)>4m_e^2.
\end{equation} We remark that the inequality above is valid  for any value of $\mathfrak{b}$ and $m$.  Observe that,  for a  dipole magnetic field $B(r)=B_* \left(r_*/r\right)^3$ with $B_*$  the field at the star surface  and $r_*$ the star's radius,   the  condition (\ref{initialrelation}) translates into 
\[
 \mathfrak{b}_*^{1/3}\gg\frac{r}{r_*}\gg b_*^{1/3},
\]where the respective field parameters at the pulsar surface are $\mathfrak{b}_*=B_*/B_c$ and $b_*=B_*/B_0$. We note, that the emission of curvature $\gamma$ quanta takes place in a vicinity of the pulsar surface [$r\gtrsim r_*$] where  the local field can exceed the characteristic scale of QED  [$b_*\gtrsim 1$]. It is worth remarking that, for pulsars holding  $b_*^{1/3}<1$, the inequality in the right hand side is bounded by  unity. Hence, instead of the relation above we should consider
\begin{equation}\label{setrrwd}
 \mathfrak{b}_*^{1/3}\gg\frac{r}{r_*}>\mathrm{max}\{1, b_*^{1/3}\}.
\end{equation}

For  ALPs  satisfying  the relations  $m\mathfrak{b}< 2m_e$ and $\mathfrak{b}\gg1$ at the moment when  $b\ll1$\footnote{These two conditions imply  that locally $ m/(2m_e)<\mathfrak{b}^{-1}\ll 1$. Hence, we are basically referring to $m\ll 1\;\mathrm{MeV}$.} the characteristic value of $q_\perp$ approximates to $q_{\perp,\mathrm{thr}}\approx 2m_e[1+m\mathfrak{b}/(2m_e)]$.   In a dipole field, the condition  $m\mathfrak{b}< 2m_e$ introduces an additional restriction in Eq. \eqref{setrrwd}:
\begin{equation}
\mathfrak{b}_*^{1/3}\gg\frac{r}{r_*}>\mathrm{max}\left\{1, b_*^{1/3},\left(\frac{m\mathfrak{b}_*}{2m_e}\right)^{1/3}\right\}.\label{zzss}
\end{equation}This means that  the fraction of  extraordinary photons  with  $q_\perp\ll 2m_e$,  having traversed a radial distance $r$ which satisfies \eqref{zzss}  will escape from the pulsar.  This effect  could explain  the observed hard $\gamma-$rays emission  $\gtrsim 10^3\; \mathrm{GeV}$ from neutron stars, and might even provide new bounds on the parameter space of ALPs   by requiring that the energy  loss  due to this mechanism does not exceed the corresponding luminosity of the pulsar.

\begin{wrapfigure}{l}{0.5\textwidth}
\centering
\includegraphics[width=0.5 \textwidth]{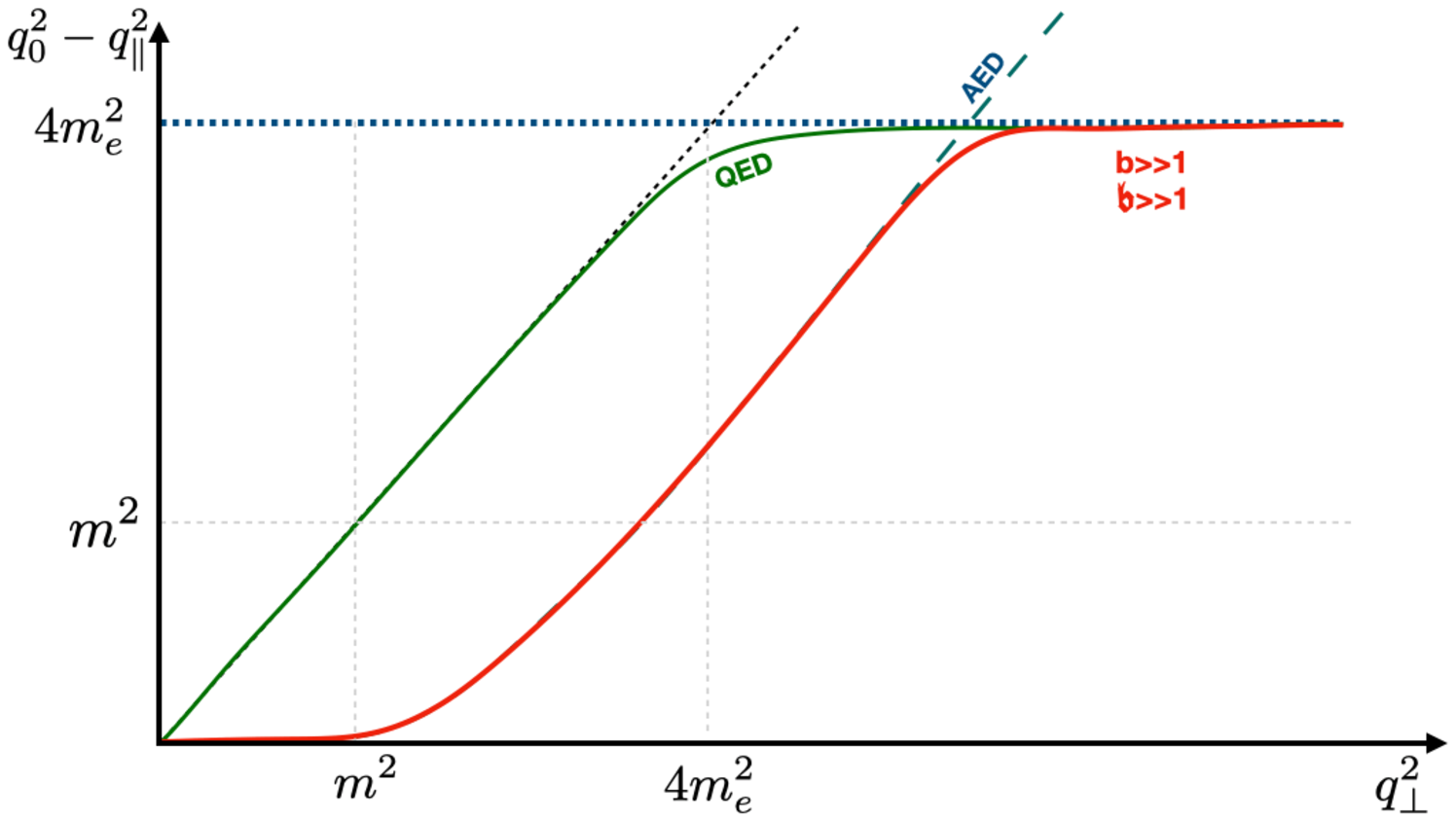} 
\caption{Sketch of the  dispersion relation for an extraordinary mode in an inhomogeneous magnetic field (red). The curve describes  the scenario in which the $\gamma$ quantum reaches $q_\perp>2m_e$ while $\mathfrak{b},\; b\gg1$.  The green curve  corresponds to the pure QED case,  whereas the dashed blue line is  linked to pure AED.  As $q_0^2-q_\parallel^2\to 4m_e^2$, the combined scenario of QED+AED  leads to a dispersion law which tends to be parallel to the dotted line that demarcates the first pair creation threshold. Once the field is such that  $b<1$ but $\mathfrak{b}\gg1$, both the green  and  red curves are expected to cross the threshold. } \label{fig:ASEp}\vspace{-10pt}
\end{wrapfigure}
On the other hand,  the fraction of extraordinary photons  with $q_\perp> 2m_e[1+m\mathfrak{b}/(2m_e)]$  that has traversed  a distance $r$  according to   \eqref{zzss}, could decay into electron-positron pairs  because their dispersion curves could cross the first threshold.  Clearly,   until  fulfilling  the aforementioned condition,  these quanta   could undergo a  magnetic field strength characterized by  $b\gg1$. If during  the time interval  in which the previous condition applies both  $q_\perp> 2m_e[1+m\mathfrak{b}/(2m_e)]$ and $q_{0,e}^2-q_\parallel^2\approx 4m_e^2$ are met,  the squared-root singularity of the  QED  polarization tensor  becomes relevant \cite{Shabad:1975ik}. Under such  circumstances, QED restores  the control of the phenomenology for a while.  This means that  an extraordinary photon which has been canalized initially  due to the strong axion-induced refraction, i.e. having $q_\perp\ll m \mathfrak{b}<2m_e$ ,could  be recaptured  later on owing to the  strong birefringent property  of the  vacuum driven by the quantum vacuum  fluctuations of the electron-positron field.  As the field decreases gradually,  the capture due to the aforementioned singularity is expected to end, and eventually,  the dispersion laws of these  photons will   cross  the first pair creation threshold.

 It is worth remarking  that, under such a circumstance,  the capture undergone by this part of the  wave  would  induce  an effective growing of the mean  free path of the curvature  radiation   as compared to the case  in which the vacuum polarization effects are considered perturbatively. This fact  constitutes a striking effect that  puts under scrutiny  pulsar models  in  which  the  curvature radiation is supposed to propagate straightforwardly and  to supply  the  electron-positron plasma \cite{Sturrock,Arons,Ruderman}. In these models the plasma  serves  to limit a required  polar gap the height of which --  relative to the star surface  --  cannot exceed considerably the mean free path  that a  curvature $\gamma-$quantum  traverses before creating an electron-positron  pair \cite{Ruderman}.  We emphasize that  the extension  of the polar gap due to the axion-mediated capture  of the extraordinary mode would modify  the so far accepted expression for the pulsar luminosity which is proportional to the square of the gap height . 

\section{Discussion and outlook}
%%%%%%%%%%%%%%%%%%%%%%%%%

Magnetic fields stronger than the characteristic scale of AED can induce nontrivial modifications  of the   QED  phenomenology. In this contribution we have explored the scenario in which these modifications prevail  hypothetically over the QED.  It has been  shown  that, in super strong  magnetic fields [$\mathfrak{b}\gg1$],  electromagnetic processes  characterized   by perpendicular momentum transfers much smaller than the axion mass tend to be suppressed due  to the dimensional reduction that the  gauge sector  responsible for the electrostatic interaction  undergoes. This effect represents  a manifestation of the anisotropy of the running  electromagnetic coupling  that  the axion-mediated  vacuum polarization  causes. 

In the second part we  evaluated   some   implications  on the $\gamma$ emission mechanism of pulsars.  This preliminary analysis indicates that, in a  strong field regime -- characterized  by the condition  $B\gg B_c$ with $B_c\ll B_0$ -- ALPs could   enable  the escape  of  certain part of the  curvature radiation.  The escaping  of these photons could explain the observed hard $\gamma-$rays emission  from pulsars.  Hence,  establishing how the gap height  and the corresponding pulsar luminosity are  modified by the  axion-diphoton interaction would open an enticing window to probe the existence of axion-like particles. This will be the subject of a forthcoming investigation.

%We have established the conditions for which the $\gamma-$radiation does not produce electron-positron pairs, essential for determining  the gap height.  The region of ALPs %parameter allowing for this total suppression of  the  plasma has been discarded by considering certain neutron stars in which  the plasma is likely that the plasma exist in concordance  %with  the polar gap model.  However, our analysis does not take into account a hypothetical interaction between ALPs  and the  Dirac field of the form $\sim\bar{\psi}\phi\psi$, which   %might benefit  the plasma formation via ALPs decay, for instance.  Hence our study assume implicitly that the phenomenology linked to this interplay  is negligible as compared with the %one treated here. 

%%%%%%%%%%%%%%%%%%%%%%%%%%%%%%%%%%
\subsection{Acknowledgments}
%%%%%%%%%%%%%%%%%%%%%%%%%%%%%%%%%%

Villalba-Ch\'avez~S and M\"uller~C  gratefully acknowledge the funding by the German Research Foundation (DFG) under Grants No.~258838303 (MU 3149/2-1) and No.~388720772 (MU 3149/5-1).  Shabad~A~E  acknowledges the support of the Russian Foundation for Basic Research under the project No. 20-02-00193.

%%%%%%%%%%%%%%%%%%%%%%%%%%%%%%%%%%
\section*{References}
%%%%%%%%%%%%%%%%%%%%%%%%%%%%%%%%%%

\end{document}